\documentclass[aps,twocolumn,prd,epsf]{revtex4}
\usepackage{graphicx}
\newcommand{\beq}{\begin{equation}}
\newcommand{\eeq}{\end{equation}}
\newcommand{\mpl}{M_{\rm pl}}
\newcommand{\lpl}{\ell_{\rm pl}}

\newcommand{\R}{{\cal R}}

\def\ba{\begin{eqnarray}}
\def\ea{\end{eqnarray}}

\begin{document}

\title{Loop quantum gravity effects on inflation and the CMB}

\author{Shinji Tsujikawa$^1$, Parampreet Singh$^2$, and
Roy Maartens$^1$}

\address{~}

\address{$^1$Institute of Cosmology and Gravitation,
University of Portsmouth, Portsmouth PO1~2EG, UK}

\affiliation{$^2$IUCAA, Ganeshkhind, Pune 411 007,
India\footnote{Current address: Institute for Gravitational
Physics \& Geometry, Pennsylvania State University, PA~16802,
USA}}

\date{\today}

\begin{abstract}

In loop quantum cosmology, the universe avoids a big bang
singularity and undergoes an early and short super-inflation
phase. During super-inflation, non-perturbative quantum
corrections to the dynamics drive an inflaton field up its
potential hill, thus setting the initial conditions for standard
inflation. We show that this effect can raise the inflaton high
enough to achieve sufficient e-foldings in the standard inflation
era. We analyze the cosmological perturbations generated when
slow-roll is violated after super-inflation, and show that loop
quantum effects can in principle leave an indirect signature on
the largest scales in the CMB, with some loss of power and running
of the spectral index.

\end{abstract}


\maketitle

\section{Introduction}

The inflationary paradigm has been very successful, solving
various problems in the big bang model of cosmology, and providing
a framework for understanding structure formation in the Universe.
Recent observational data on the cosmic microwave background (CMB)
anisotropies, together with other data, are well accounted for by
a simple inflationary model with suitable cold dark matter and
dark energy content~\cite{Spergel}.

However, this simple picture belies a number of fundamental
puzzles. In particular, there is the problem of how to explain the
initial conditions for successful inflation; for example, with the
simple potential,
 \beq \label{po}
V(\phi)={1\over2} m_{\phi}^2\phi^2\,,
 \eeq
one requires an initial inflaton value $\phi_i\gtrsim 3\mpl$. The
``eternal inflation" scenario provides an answer to this, relying
on the infinite extent of space, but its arguments have been
challenged~\cite{hawking}. In any event, whatever the merits of
eternal inflation, it is useful to seek alternative explanations
and to explore what quantum gravity may tell us about this.
Quantum gravity effects above the Planck energy $\mpl$ should be
able to set the initial conditions for inflation--or to provide an
alternative to inflation with adequate predictive power for
structure formation. This is one of the challenges facing
candidate theories of quantum gravity.

Results from WMAP also point to a possible loss of power at the
largest scales and running of the spectral index. These effects,
if confirmed, are difficult to explain within standard slow-roll
single-field inflation, and may have roots in Planck-scale
physics.

Loop quantum gravity (or quantum geometry) is a candidate quantum
gravity theory that has recently been applied to early universe
cosmology (see~\cite{Bojo_review} for a recent review). It
improves on a conventional Wheeler-DeWitt quantization, by
discretizing spacetime and thus avoiding the breakdown of the
quantum evolution even when the classical volume becomes
zero~\cite{Bojo1}.

The quantization procedure involves a fiducial cell corresponding
to a fiducial metric to define a symplectic structure.
Introduction of a fiducial cell, which is not required at a
classical level, is a necessity for Hamiltonian quantization. This
procedure has been developed for spatially flat and closed
Friedmann geometries (and related Bianchi geometries). It is
important to note that the scale factor in the quantum regime has
a different meaning than in classical general
relativity~\cite{thanks}. For a flat geometry (which is the case
we consider), in the quantum regime the scale factor is not
subject to arbitrary rescaling, as in general relativity. Instead,
the scale factor $\tilde a$, obtained from loop quantization by
redefining canonical variables, is related to the conventional
scale factor $a$ via
\begin{equation}
\tilde {a}^3 = a^3 V_0\,,
\end{equation}
where $V_0$ is the volume of the fiducial cell. In this definition
the quantization is independent of the rescaling of the fiducial
metric and thus invariant under rescaling of the conventional
scale factor. However, one can also quantize without redefining
the canonical variables. In this case it has been shown that
states in Hilbert space corresponding to two different fiducial
metrics are unitarily related and the physics does not change with
rescaling of the fiducial metric~\cite{abl}. This is consistent
with the background-independence of loop quantization.

What emerges in the quantization is a fundamental length scale,
\begin{equation}\label{ls}
\tilde{a}_*=\ell_*\equiv\sqrt{\gamma j \over 3}\,\lpl\,,
\end{equation}
where $\gamma \, (\approx 0.13)$ is the Barbero-Immirzi parameter
and $j $ is a half-integer (where $j>1$ is necessary in order to
study evolution via an effective matter Hamiltonian~\cite{Bojo2}).
This length scale corresponds to a particular effective quantum
volume (or the state of the universe) below which the dynamics of
the universe is significantly modified by loop quantum effects.
The crucial quantity that determines the dynamics and quantifies
the state of the universe relative to the critical state, is
\begin{equation}
q\equiv {\tilde {a}^2 \over \tilde {a}_*^2} = {a^2 \over a_*^2}\,.
\end{equation}

The Planck length $\lpl$ is not put in by hand, but arises from
the quantization procedure. The fundamental scale $\ell_*$ is the
same for a flat or closed classical geometry and has nothing to do
with the topology. Thus for a flat non-compact spacetime, there is
a fundamental length scale defined by quantization, unlike general
relativity, where no length scale (other than the Hubble length)
is defined by the geometry. If a compact topology is imposed on a
flat geometry, then that introduces another (classical) length
scale, independent of $\ell_*$. Here we consider a flat
non-compact universe. In the semi-classical regime, where
spacetime is well approximated as a continuum but the dynamics is
subject to non-perturbative loop quantum corrections, the
conventional scale factor has the usual rescaling freedom. For
convenience, we use the rescaling freedom to fix the classical
scale factor at the critical epoch of transition from quantum to
classical evolution:
\begin{equation}
a_*=\ell_*\,.
\end{equation}
It should be noted that any other choice could be used, or the
rescaling freedom could be kept. The point is that the relevant
physical quantity $q$ remains invariant.

\section{Non-perturbative semi-classical dynamics}

In loop quantization, the geometrical density operator has
eigenvalues~\cite{abl,Bojo2}
 \beq \label{dj}
d_j(a)=D(q)\,{1\over a^3}\,,
 \eeq
where the quantum correction factor for the density in the
semi-classical regime is
 \ba \label{p}
&&D(q) = \left( {8\over 77}\right)^6 q^{3/2} \Big\{7
\Big[(q+1)^{11/4}
-|q-1|^{11/4}\Big] \nonumber \\
&&~~~{}- 11q\Big[(q+1)^{7/4}-{\rm sgn}\,(q-1) |q-1|^{7/4}\Big]
\Big\}^6\!.
 \ea
In the classical limit we recover the expected behaviour of the
density, while the quantum regime shows a radical departure from
classical behaviour:
 \ba\label{dc}
a \gg a_* &\Rightarrow & D \approx 1\,,\\
\label{dq} a \ll a_* & \Rightarrow & D \approx
\left({12\over7}\right)^6\left({a \over a_*}\right)^{15}\,.
 \ea
Then $d_j$ remains finite as $a \to 0$, unlike in conventional
quantum cosmology, thus evading the problem of the big-bang
singularity. Intuitively, one can think of the modified behaviour
as meaning that classical gravity, which is always attractive,
becomes repulsive at small scales when quantized. This effect can
produce a bounce where classically there would be a singularity,
and can also provide a new mechanism for inflationary
acceleration.

In loop quantum cosmology, a scalar field $\phi$ with potential
$V(\phi)$ in a flat Friedmann-Robertson-Walker background is
described by the Hamiltonian~\cite{Bojo2}
 \beq \label{hamiltonian}
{\cal H}= a^3 V(\phi)+ {1\over2}d_j \, {p_\phi^2}\,,~ p_\phi =
d_j^{-1} \dot \phi\,,
 \eeq
where $p_\phi$ is the momentum canonically conjugate to $\phi$.
This gives rise to an effective Friedmann equation for the Hubble
rate $H=\dot{a}/a$,
 \beq \label{back}
H^2 = {8\pi \lpl^2\over 3}\left[V(\phi) + {1\over 2D} \dot{\phi}^2
\right],
 \eeq
together with the modified Klein-Gordon equation
 \beq
\ddot{\phi}+ \left(3H- {\dot{D}\over D}\right) \dot{\phi}+D
V_\phi(\phi)=0 \,. \label{backp}
 \eeq
The quantum corrected Raychaudhuri equation follows from
Eqs.~(\ref{back}) and (\ref{backp}),
 \beq \label{dotH}
\dot{H}=-{4\pi}{\lpl^2} \dot{\phi}^2\,{1\over D}
\left(1-{\dot{D}\over 6HD}\right)\,.
\eeq

In the quantum limit, Eq.~(\ref{dq}) shows that the kinetic terms
dominate the potential terms in Eqs.~(\ref{back}) and
(\ref{backp}), and this leads to:
 \beq
\dot{\phi} \propto a^{12}\,,~~a \propto (-\eta)^{-2/11}\,,
 \eeq
where $\eta$ is conformal time. Since $\dot{D}/(HD)=15$ for $a \ll
a_*$, we have from Eq.~(\ref{dotH})
\begin{equation}
\dot{H} \approx 6\pi\lpl^2 {\dot{\phi}^2 \over D}>0\,.
\end{equation}
Quantum effects thus drive super-inflationary
expansion~\cite{Bojo2}. However, this can not yield sufficient
e-folds to overcome the flatness and horizon problems in the
absence of the inflaton potential.

In the quantum regime, $a \ll a_*$, when $d_j$ behaves as a
positive power of $a$, the second term on the left of
Eq.~(\ref{backp}) acts like an anti-friction term and pushes the
inflaton up the hill. The strong dominance of the kinetic term
over the potential term means not only that the mechanism is
robust to a change in $V(\phi)$, but also that it dominates over
gradient terms in $\phi$.

\section{Inflation with loop quantum modifications}

With the growth in $a$, the eigenvalue $d_j$ eventually behaves as
in Eq.~(\ref{dc}). The second term on the left of
Eq.~(\ref{backp}) then behaves as a friction term, but it takes
some time before this term halts the motion of $\phi$, since the
initial quantum push is very strong. Then the field begins to roll
down the potential from its maximum value $\phi_{\rm max}$,
initiating a standard slow-roll inflationary stage.

We assume the inflaton is initially at the minimum of its
potential. Initial small quantum fluctuations are sufficient to
start the process of raising the inflaton up the hill. Strong
kinetic effects will rapidly overwhelm the quantum fluctuations.
We assume these fluctuations are constrained by the uncertainty
principle, $|\Delta\phi_i \Delta p_{\phi\,i}|>1$, so that
 \beq \label{up}
|\phi_i\dot\phi_i| \gtrsim {10^3\over j^{3/2}} \left({a_i \over
a_*}\right)^{12} \mpl^3\,.
 \eeq
We take $a_i = \sqrt{\gamma} l_{pl}$ (for $a_i\ll \lpl$, space is
discretized and we cannot use the smooth dynamical equations
above). The sign of $\dot \phi_i$ determines whether the inflaton
just moves further up the hill (for $\dot \phi_i> 0$) or returns
to $\phi = 0$ and then is pushed up (for $\dot \phi_i < 0$).

\begin{figure}
\begin{center}
\includegraphics[width=9cm,height=7.5cm]{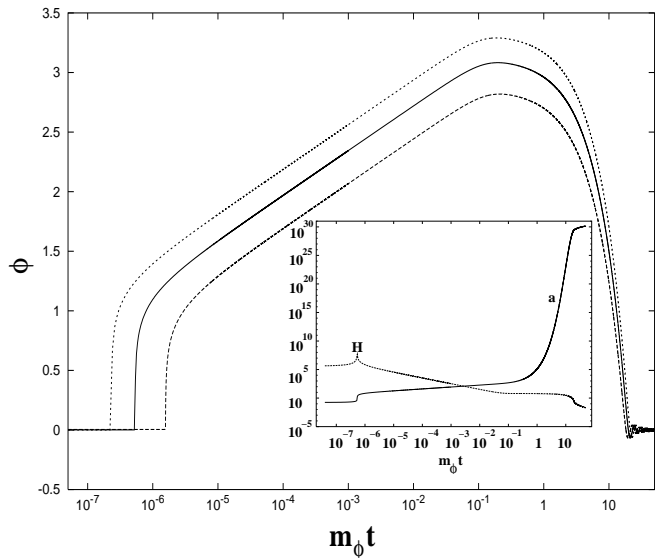}
\end{center}
\caption{Evolution of the inflaton (in Planck units). We set
$\dot{\phi}_{i}/(m_\phi\mpl)=2$, with $m_\phi/\mpl=10^{-6}$, and
choose $\phi_i/\mpl$ as the minimum value satisfying the
uncertainty bound, Eq.~(\ref{up}), i.e. $\phi_i/\mpl \sim
10^{12}j^{-15/2}$. The solid curve has $j = 100$, the upper dashed
curve has $j = 125$, and the lower dashed curve has $j = 75$.\\
{\bf Inset}: Evolution of the scale factor and Hubble rate (in
units of $m_\phi$) with the same parameters as the solid curve for
$\phi$. } \label{evolution}
\end{figure}

\begin{figure}
\begin{center}
\includegraphics[width=9cm,height=7.5cm]{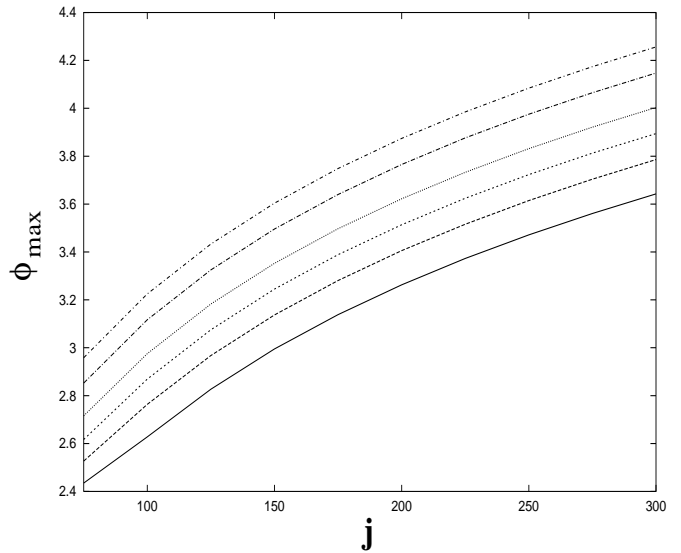}
\end{center}
\caption{The maximum reached by $\phi$ (in Planck units) as a
function of $j$. From top to bottom, the curves correspond to
initial conditions $\dot{\phi}_i/(m_\phi\mpl) = 5, 2.5, 1, 0.5,
0.25, 0.1$ and $\phi_i/\mpl$ is taken as the minimum value
satisfying the uncertainty bound, Eq.~(\ref{up}). An increase
(decrease) in $j$ and ${\dot \phi_i}$ leads to an increase
(decrease) in the number of e-folds. } \label{j_phimax}
\end{figure}

In order to make this qualitative description more precise,
consider the simple potential, Eq.~(\ref{po}). Large-scale CMB
anisotropies require that~\cite{LL}
\begin{equation}
\phi_{\rm max} \gtrsim 3\mpl\,,~~~m_\phi\sim 10^{-6}\mpl\,.
\end{equation}
We find that loop quantum effects can produce a large enough
$\phi_{\rm max}$ even for very small $\phi_i/\mpl$ and
$\dot\phi_i/\mpl^2$ [satisfying the uncertainty constraint,
Eq.~(\ref{up})]. The results are shown in Figs.~\ref{evolution}
and \ref{j_phimax}.

In the quantum regime ($a < a_*$), the energy density of the
universe is dominated by the kinetic energy of the inflaton. Since
$D^{-1} \sim q^{-15/2} \gg 1$ for $q \ll 1$ in Eq.~(\ref{back}),
the Hubble rate can take large values even if $\dot{\phi}^2$ is
small. As seen in the inset of Fig.~\ref{evolution}, the Hubble
rate increases for $a \lesssim a_*$ due to the growth of the
kinetic term in Eq.~(\ref{back}).

By  Eq.~(\ref{dotH}), super-inflation ends when $\dot{D}/(6HD)=1$.
We find numerically that this term quickly goes to zero just after
the Hubble maximum. Figure~\ref{evolution} also shows the short
non-inflationary phase ($\ddot a<0$) while the field continues
rolling up. The second stage of inflation begins as the inflaton
approaches its maximum ($\dot{\phi}=0$). Since $d_{j}(a) \approx
a^{-3}$ at this stage, this is standard chaotic inflation,
followed by conventional reheating when the inflaton oscillates.

In Fig.~\ref{j_phimax} we plot the maximum value $\phi_{\rm max}$
reached via quantum gravity effects, for various values of $j$ and
$\dot{\phi}_i$. Sufficient inflation ($\gtrsim 60$ e-folds)
requires $\phi_{\rm max} \gtrsim 3M_{\rm pl}$ and
Fig.~\ref{j_phimax} shows this is possible for a wide range of
parameters. An increase in $j$ decreases $q$ and $d_j(a)$, thereby
yielding larger $H$ and $\ddot{\phi}$, and so increasing the
number of e-folds.

Loop quantum effects can therefore in principle set the initial
conditions for successful slow-roll inflation. A further question
is whether any observational signature of the first phase of
quantum gravity inflation survives the second phase of classical
inflation. It should be noted that violation of slow roll occurs
{\em after} the super-inflationary phase has ended and the
universe is classical. In this sense, violation of slow roll is an
{\em indirect} loop quantum gravity effect. The violation of the
slow-roll condition around $\dot{\phi}=0$, which is peculiar to
this scenario, can lead to some suppression of the power spectrum
at large scales and running of the spectral index, provided that
\begin{equation}\label{phil}
\phi_{\rm max}\sim \phi_{\rm ls}\,,
\end{equation}
where $t_{\rm ls}$ is the time when the largest cosmological
scales are generated. If $\phi_{\rm max}\gg \phi_{\rm ls}$, i.e.
if quantum gravity effects drive the inflaton far up the hill,
then cosmological scales are generated well into the classical era
and there is {\em no} loop quantum signature in the currently
observed CMB.

\section{CMB anisotropies}

To be more concrete, we consider cosmological perturbations
generated when the universe is in the classical regime ($a \gg
a_*$), but slow-roll is violated. The spectrum of comoving
curvature perturbations, ${\cal R}$, generated in slow-roll
inflation is given by~\cite{LL}
 \ba
&& {\cal P}_{\R} \propto k^{n-1}\,,~~n \approx 1-6\epsilon+2\eta
\,,\\ && \epsilon \equiv \frac{\mpl^2}
{16\pi}\left(\frac{V_\phi}{V}\right)^2\,,~~\eta \equiv
\frac{\mpl^2}{8\pi}\frac{V_{\phi \phi}}{V}\,.
 \ea
For the potential~(\ref{po}), this yields a slightly red-tilted
spectrum,
 \beq
n \approx 1-\frac{1}{\pi}\left(\frac{\mpl}{\phi}\right)^2 \,.
\label{PR2}
 \eeq

The amplitude of the power spectrum generated in slow-roll
inflation is~\cite{LL} ${\cal P}_{\R} \approx H^4/\dot{\phi}^2$.
In the loop quantum scenario, $\phi$ changes direction after it
climbs up the potential hill to its maximum, $\dot \phi = 0$. It
was shown in Ref.~\cite{Seto} that the spectrum of curvature
perturbations is not divergent even at $\dot{\phi}=0$, when the
standard formula breaks down and should be replaced by
\begin{eqnarray}
{\cal P}_{\R} \approx \frac{9H^6}{V_{\phi}^2} \approx
\frac{9H^6}{m_\phi^4\phi^2}\,. \label{calPR2}
\end{eqnarray}
This appears to correspond to the replacement of $\dot{\phi}$ in
the standard formula by the slow-roll velocity $\dot{\phi}
=-V_{\phi}/(3H)$, but Eq.~(\ref{calPR2}) is valid even at
$\dot{\phi}=0$, when the slow-roll approximation breaks
down~\cite{Seto}.

Using Eq.~(\ref{calPR2}), the spectral index becomes
\begin{eqnarray}
n=1+\frac{1}{1+\epsilon_1}
\left(6\epsilon_1-\frac{2\dot{\phi}}{H\phi}\right)\,,
\label{nRmodify}
\end{eqnarray}
where $\epsilon_{1} \equiv {\dot H}/{H^2}$, which vanishes at
$\dot \phi = 0$ and thus leads to a scale-invariant spectrum at
the turn-over of the inflaton. This modification of the spectral
index around $\dot{\phi}=0$ is a distinct feature of the loop
quantum scenario. The formula~(\ref{nRmodify}) reduces to the
standard one in the slow-roll region ($|\ddot{\phi}| \ll
|3H\dot{\phi}|$), since $\dot{\phi} =-V_{\phi}/(3H)$ and
$|\epsilon_1| \ll 1$.

When the field climbs up the potential hill ($\dot{\phi}>0$), we
have $n<1$ by Eq.~(\ref{nRmodify}). Therefore the spectrum begins
to grow toward large scales for modes which exit the Hubble radius
during the transient regime. However the power spectrum is close
to scale-invariant as long as $\dot{\phi}$ is close to zero. We
should also mention that there exists a short non-inflationary
phase after the Hubble maximum (see Fig.~\ref{evolution}). However
the fluctuations in this non-inflationary phase are not of
relevance, since the standard causal mechanism for the generation
of perturbations does not operate.

The spectral index can be expanded as
\begin{eqnarray}
n(k) &=& n(k_0)+ {\alpha (k_{0})\over 2}\, \ln \left({k \over
k_0}\right) + \cdots\,, \label{sperunning}\\ \alpha &=& \left(
{{\rm d}n \over {\rm d}\ln k} \right)_{k=aH}~,
\end{eqnarray}
where $k_0$ is some pivot wavenumber. In our case, the spectral
index changes rapidly around the region $\dot{\phi}=0$, which
leads to larger running compared to the slow-roll regime. This
means that it is not a good approximation to use a constant
running in the whole range of the potential-driven inflation. For
example one has $n \sim 0.964$ and $\alpha \sim -0.005$ at
$\phi=3M_{\rm pl}$ from Eq.~(\ref{PR2}). We find numerically that
the running becomes stronger around $\dot{\phi}=0$, with minimum
value $\alpha \sim -0.06$.

\begin{figure}
\begin{center}
\includegraphics[width=9cm,height=7cm]{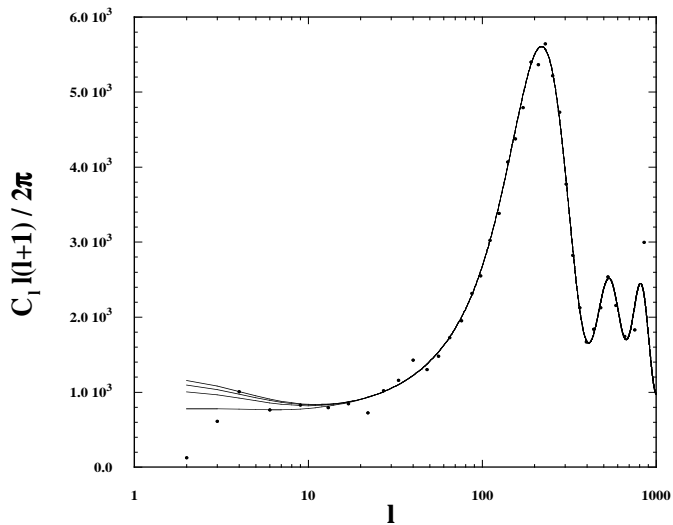}
\end{center}
\caption{The CMB angular power spectrum with loop quantum
inflation effects. From top to bottom, the curves correspond to
(i)~no loop quantum era (standard slow-roll chaotic inflation),
(ii)~$\bar{\alpha}=-0.04$ for $k \le k_0=2 \times 10^{-3}\,{\rm
Mpc}^{-1}$, (iii)~$\bar{\alpha}=-0.1$ for $k \le k_0$, and
(iv)~$\bar{\alpha}=-0.3$ for $k \le k_0$. There is some
suppression of power on large scales due to the running of the
spectral index. } \label{Cl}
\end{figure}

Consider a scenario in which the spectral index grows rapidly
towards 1 around $\phi=3M_{\rm pl}$, corresponding to the pivot
scale $k_0 \sim 10^{-3}\,{\rm Mpc}^{-1}$. One can express $n$ in
this region in terms of the average value $\bar\alpha$, i.e.,
$n(k) \approx n(k_0)+ (\bar{\alpha}/2){\ln}(k/k_0)$. Then we
obtain $n(k) \sim 1$ at $k \sim 0.1k_0$ for $\bar{\alpha}=-0.04$.
This behavior was found numerically, and leads to a larger
spectral index around the scale $k \sim 10^{-4}\,{\rm Mpc}^{-1}$,
compared to the standard slow-roll chaotic inflationary scenario.

In Fig.~\ref{Cl} the CMB power spectrum is plotted for several
values of $\bar{\alpha}$ for $k \le k_0$, and with the average
running $\bar{\alpha}=-0.005$ for $k>k_0$. In standard chaotic
inflation $|\alpha|\lesssim 0.01$ around $\phi \sim 3M_{\rm pl}$,
in which case it is difficult to explain the running with
$\alpha=-0.077^{+0.050}_{-0.052}$ around the scale $k \sim 10^{-3}
\,{\rm Mpc}^{-1}$ reported by the WMAP team~\cite{Spergel}. In the
loop quantum inflation scenario, it is possible to explain this
running of the power spectrum due to the existence of the
non-slow-roll region ($\dot{\phi} \sim 0$) that follows loop
quantum inflation--provided that Eq.~(\ref{phil}) holds, i.e., the
loop quantum inflation does not push the inflaton too high up its
potential hill. If the e-folds in slow-roll inflation are greater
than 60, then the CMB power spectrum does not carry a loop quantum
signature~\cite{bf}. Although strong suppression of power around
the multipoles $l=2, 3$ is difficult to obtain unless
$\bar{\alpha} \lesssim -0.3$, it is quite intriguing that the loop
quantum scenario can provide a possible way to explain the
observationally supported running of the spectrum, with some loss
of power on large scales, albeit with fine-tuning of parameters.

\section{Conclusions}

In summary, we have shown that loop quantum effects can drive the
inflaton from equilibrium up its potential hill, and can thus in
principle set appropriate initial conditions for successful
standard inflation. Loop effects may also leave an indirect
imprint on the CMB on the largest scales, if the inflaton is not
driven too far up its potential hill. This is possible in
particular for very small initial fluctuations in the inflaton and
its velocity, compatible with the uncertainty principle,
Eq.~(\ref{up}). We stress that the observational signature we
found is an indirect effect of loop quantum gravity, rather than a
direct consequence of perturbations generated in the regime
$a<a_{*}$. In the loop quantum super-inflationary era, the
standard quantization using the Bunch-Davies vacuum is no longer
valid. Further analysis is required to understand the quantization
of metric perturbations in loop quantum gravity.

By analyzing the dynamics and perturbations in the non-slow roll
regime that follows the super-inflation, we showed that the loop
quantum effects can indirectly lead to a running of the spectral
index, with some loss of power on the largest scales. It should be
noted that a similar loss of power on the largest scales can also
be obtained via string theoretic scenarios (see Ref.~\cite{other}
and references therein). It would be interesting to distinguish
between the different cosmological effects predicted by loop
quantum gravity and string theories, using future high-precision
observations.

\vspace*{0.2cm}
\noindent{\bf Acknowledgements}\, We thank A. Ashtekar,
B. A. Bassett, M. Bojowald, N. Dadhich, D. Jatkar and M. Sami for very
helpful discussions. S.T. is supported by JSPS (No.~04942), P.S.
by CSIR, R.M. by PPARC. P.S. thanks the ICG Portsmouth for support
for a visit during which this work was initiated.

\end{document}